\documentclass[pdftex,12pt]{JHEP3}
\usepackage{amsmath,amsfonts,amsbsy,latexsym,amssymb,amscd,amstext}
\usepackage[pdftex]{epsfig}

\pdfoutput=1

 \def\be{\begin{equation}}
\def\ee{\end{equation}}
\def\bea{\begin{eqnarray}}
\def\eea{\end{eqnarray}}
\def\pd{\partial}
\def\a{\alpha}
\def\b{\beta}

\def\d{\delta}
\def\m{\mu}
\def\n{\nu}

\def\l{\lambda}

\def\r{\rho}
\def\s{\sigma}
\def\e{\epsilon}

\def\bi{\begin{itemize}}
\def\ei{\end{itemize}}

\date{May 25th, 2008} \preprint{IFT-UAM/CSIC-09-23}
\title{Transverse gravity versus observations} \author{Enrique \'Alvarez, Ant\'on F. Faedo and J.J. L\'opez-Villarejo $^a$ \\ $^a$ Instituto de F\'{\i}sica Te\'orica
UAM/CSIC and Departamento de F\'{\i}sica Te\'orica \\ Universidad
Aut\'onoma de Madrid, E-28049--Madrid, Spain \\ E-mail: \email{enrique.alvarez@uam.es }}
\abstract{Theories of gravity invariant under those diffeomorphisms generated by transverse vectors, $\pd_\m\xi^\m=0$ are considered. Such theories are dubbed transverse, and differ from General Relativity in that the determinant of the metric, $g$, is a transverse scalar. We comment on diverse ways in which these models can be constrained using a variety of observations. Generically, an additional scalar degree of freedom mediates the interaction, so the usual constraints on scalar-tensor theories have to be imposed. If the purely gravitational part is Einstein--Hilbert but the matter action is transverse, the models predict that the three {\em a priori} different concepts of mass (gravitational active and gravitational passive as well as inertial)  are not equivalent anymore. These transverse deviations from General Relativity are therefore tightly constrained, actually correlated with existing bounds on violations of the equivalence principle, local violations of Newton's third law and/or violation of Local Position Invariance.}
\begin{document}

{\vskip 1cm}

\section{Introduction}

This paper is devoted to study some observational consequences of the hypothesis that the set of symmetries enjoyed by the theory that describes gravity is not the full group of diffeomorphisms (Diff(M)), as in General Relativity, but a maximal subgroup of it. Transformations belonging to this subgroup have been dubbed {\em transverse} \cite{Alvarez} (TDiff(M)) since at the infinitesimal level the parameter describing the coordinate change $x^\m\to x^\m+\xi^\m(x)$ is transverse, i.e., $\pd_\m\xi^\m=0$. Incidentally, this is the smaller symmetry one needs to propagate consistently a graviton \cite{vanderBij}, which is a great theoretical motivation for considering these theories. Other theoretical aspects, including the coupling to matter and ultraviolet properties have been considered in \cite{Alvarez, Alvarezfaedo, Comment, UV}.
\par
At the non-linear level, probably the best way to understand transverse theories is as those invariant under the subgroup of diffeomorphisms that preserve some measure. At the end, this restriction forces the determinant of the Jacobian of the transformation to equal unity. The most important consequence is that we can no longer distinguish between tensor densities and true tensors. In particular, the determinant of the metric is a true scalar and thus the symmetry does not fix the function dressing neither the Ricci scalar nor the matter Lagrangian to be the square root of the determinant.     
\par
On the other hand, one of the distinctive features of the lack of symmetry is the impossibility to avoid propagation of an additional scalar mode contained in the metric besides the usual spin two graviton. This means that, generically, gravity will be mediated also by the scalar mode, much like in ordinary scalar-tensor theories or even in $f(R)$ gravity. Therefore, a straightforward path to constrain transverse models is to use the vast literature devoted to find such a scalar companion of the graviton. A variety of observations and experiments have failed to encounter this mode. Of course one can always postulate mechanisms to hide it, but there are constraints on the form of its couplings to matter, as well as its selfcoupling, coming for example from Solar System tests, deviations from Newton's law, cosmological evolution, Binary Pulsars etc.
\par
Nevertheless, first steps along this direction were taken in \cite{Comment}, where the parameters defining the transverse model at the linear level, including a possible mass for the scalar mode, were bounded using the negative results on deviations from Newton's law in the form of a Yukawa potential. The fact that the symmetry group is now smaller means that more arbitrary functions are allowed and have to be constrained by experiments. The most general quadratic Lagrangian invariant under transverse diffeomorphisms reads\footnote{The function $f(g)$ needs to be positive for the gravitons to carry positive energy. Similarly, it must be verified
\be
2(n-1)f^\prime-(n-2)f^2 f_k \geq 0
\ee
to avoid ghostly excitations of the scalar mode. The term $f_\l(g)\Lambda$ is both a potential term for the determinant and a generalization of a cosmological constant.} 
\be
S=-\frac{1}{2\kappa^2}\int\,d^nx\,\sqrt{g}\left[f(g)R+2f_\l(g)\Lambda+\frac{1}{2}f_k(g)g^{\m\n}\pd_\m g\pd_\n g\right]+S_m
\ee
where the matter action may be taken to be of the form
\be
S_m=\int\,d^nx\,f_m(g)\,L_m\left[\psi_m,g_{\m\n},g\right]
\ee
The matter Lagrangian is a functional of the matter fields $\psi_m$, the metric and its determinant. Interaction terms between matter and the determinant of the metric mean that the coupling is not minimal and in general there will be violations of the Weak Equivalence Principle (WEP). We will consider these possible violations in detail later on, but for the moment let us suppose
\bea
f_m&=&\sqrt{g}\nonumber\\
L_m\left[\psi_m,g_{\m\n},g\right]&=&L_m\left[\psi_m,g_{\m\n}\right]
\eea
Then the WEP is automatically satisfied and if we further consider the potential term to be negligible $f_\l\sim 0$ we can directly use the results of \cite{Damour} constraining the form of the coupling of the scalar mode to matter. It can be shown that the Post Newtonian parameters are given in terms of the coefficients of the expansion of the coupling function around a cosmologically imposed (and therefore evolving with time) value of the scalar mode
\bea
\gamma-1&=&\frac{4f'^2}{\frac{(n-2)^2}{2}f\,f_k+\left[(n-1)(n-2)+2\right]f'^2}\Bigg{|}_{g_0}\nonumber\\
\b-1&=&\pm\frac{1}{4(n-2)}\,\frac{f\,f'}{\frac{n-1}{n-2}f'^2+\frac{1}{2}f\,f_k}\,\frac{d\gamma}{dg}\Bigg{|}_{g_0}
\eea
The best current limits on these parameters (at the $68\%$ confidence level) are \cite{Amsler}
\bea
\gamma-1&=& (2.1 \pm 2.3) \times 10^{-5}\nonumber\\
4\b-\gamma-3&=& (4.4 \pm 4.5) \times 10^{-4}
\eea
obtained respectively from the additional Doppler shift experienced by radio-wave beams connecting the Earth to the Cassini spacecraft when they passed near the Sun and from Lunar Laser Ranging measurements of a possible polarization of the Moon toward the Sun. 
\par
Even if these limits are very restrictive in the sense that they leave little room beyond the general relativistic values $\gamma=\b=1$, it is also true that they constrain only a combination of the original functions evaluated at a point in the history of the universe. Moreover, once we drop the requirement for a vanishing potential (for example giving a mass to the scalar), constraining the form of the functions becomes even more complicated, though some simple cases are treatable using cosmological data. 
\par
In summary, what we are trying to emphasize is that models with such a large number of arbitrary functions to determine are very difficult to constrain efficiently, since usually we have enough freedom to evade the bounds. Furthermore, designing by hand the functions in such a way that every existing constraint is satisfied is not a very interesting procedure, at least from an aesthetical point of view.
\par
So, in the following, we will concentrate on some simple models that capture the features of the transversality condition and, besides, are interesting for the phenomenology of the Cosmological Constant problem. They are characterized by the absence of propagation of the additional scalar. Also, the models considered are a natural scenario for studying the consequences of a violation of some of the Principles of General Relativity. The question of the role that these theories play in the scheme of {\em metric theories} of gravity is a relevant one. We will be focused essentially on deriving observational signatures and sketching possible ways to detect them. 

\section{The Principle of Equivalence}

The {\em Weak Equivalence Principle} appears in the literature formulated in several different manners and it is not always straightforward to see why each one implies the others or, more precisely, if there is a unique way to implement them at an operational level, that is, in an action principle. So, we will take here the practical point of view of defining the WEP as the equality of {\em inertial} ($m_i$) and {\em passive gravitational} ($m_p$) masses. Inertial mass is a property of a particle independent of the environment and is characteristic in all its interactions, like other parameters (charge, gravitational mass etc.). By passive gravitational mass we mean a kind of ``charge" of the gravitational interaction, i.e., how the particle responds to an externally given gravitational field. 
\par
That both masses coincide, implying universality of the acceleration of free fall, is one of the best established experimental facts in physics, with a relative precision of at least $10^{-12}$ as quoted in the Particle Data Group \cite{Amsler}.
\par
On the other hand, the equality of the {\em active gravitational mass} ($m_a$, meaning how much gravitational field a particle generates), to the other two lies essentially on the third of Newton's laws, that is, momentum conservation\footnote{A very clear discussion about the operational significance of the various masses appears in the (extremely difficult to find) course by Deser \cite{Deser}. Notice that if we choose units in which $m_i=m_p$, the ratio between $m_p$ and $m_a$ is precisely what we call Newton's constant $G$.}, and the experimental precision seems even better. It has indeed been recently claimed by Nordtvedt  \cite{Nordtvedt} that Lunar Laser Ranging implies a bound on relative violations of Newton's third law of $10^{-13}$. Inequality between passive and active gravitational masses is traduced in an unbalanced force that accelerates the center of mass of the interacting pair
\be
\vec{F}_{12}=S(1,2)\,G\,\,m_p^1m_p^2\,\,\frac{\vec{r}_{12}}{r_{12}^3}
\ee 
To be specific, what is bound to be small is the {\em difference} of the quotient of the active and passive gravitational masses for distinct bodies, dubbed 1 and 2, that is
\be
S(1,2)=\left|\left(\frac{m_a}{m_p}\right)_1-\left(\frac{m_a}{m_p}\right)_2\right|\leq 10^{-13}
\ee
\par
After this detour regarding the WEP, let us turn our attention to transverse models and some of their implications. One of the avenues that could be explored in order to understand why the observed cosmological constant is so small (contradicting all the effective field theory wisdom so painfully accumulated over the years) is to consider that its value is the one calculated from the Standard Model but nevertheless it does not generate the gravitatory field expected by General Relativity. In other words, for vacuum energy  
\be
m_a\neq m_p
\ee
Recently, experimental bounds on this violation have been put forward in \cite{Masso}.
\par
We have explicitly shown in a previous work \cite{Alvarezfaedo} that there is an extreme transverse case for which the active mass is as small as it can be, insofar as the potential energy does not weigh \footnote{
This is not strictly true, because there is an integrability constraint relating kinetic and potential energies in these models.}. The model is
\be
S=\int d^n x\left(-\frac{1}{2\kappa^2}\sqrt{|g|}R+\frac{1}{2}g^{\m\n}\pd_\m\psi \pd_\n\psi-V\left(\psi\right)\right)
\ee
It is indeed clear why the potential energy does not weigh: it is not coupled at all to the gravitational field.
We shall show momentarily that in the geometrical optics limit (eikonal) the inertial mass corresponding to the $\psi$-particle is just the parameter appearing in the potential $m_i=m$. There is a violation of the equality between different masses in this model.

\section{Masses in transverse theories}
Let us examine the consequences of the hypothesis that the matter action is only invariant under transverse diffeomorphisms
(TDiff). For concreteness, we can bear in mind the family of models we have already pinpointed, defined by
\be\label{scalaraction}
S_m=\int d^n x f_m(g)\sum_i\left(g^{\m\n}\pd_\m\psi_i\pd_\n \psi_i-V(\psi_i)\right)\equiv \int d^n x f_m(g)L_m
\ee
where $g\equiv \det\,g_{\a\b}$ and General Relativity (GR) corresponds to $f_m=\sqrt{|g|}$. The {\em matter} content of spacetime is then represented by the fields $\psi_i$, $i=1\ldots N$. The gravitational action is taken to be the usual Einstein--Hilbert one since then we are sure to get the same solutions in vacuum. Moreover, the scalar degree of freedom mentioned in the introduction does not propagate and we do not have to take care of scalar mediation constraints. 
\par
The reason why we reduce our attention to this family of models is because then all ``particles" fall along geodesics of the metric $g_{\a\b}$. Therefore, the universality of free fall in the particle approximation is automatically fulfilled, in exactly the same sense as in GR. In order to see this, one has to remember that fields behave as particles in what is known as the geometrical optics or eikonal approximation, which consists in keeping only the dominant term in the WKB expansion of the equation of motion. Its importance stems from the fact that this is the only known way that (classical) particle behavior is obtained out of a field theory \cite{AlvarezConde}.
\par
Let us write down the equations for the physical optics approximation. Starting from the ordinary Klein-Gordon equation
\be
\left(\Box+\hat{m}^2\right)\psi=0
\ee
by expanding the field in terms of the eikonal
\be
\psi={\rm Re}\left(e^{i\left(\frac{1}{\e}\psi_0+\psi_1+\ldots\right)}\right)
\ee
if we write $\hat{m}^2\equiv \frac{m^2}{\e^2}$ then the dominant order in formal powers of $\e$ is $O(1/\e^2)$ and reads
\be
g^{\m\n}k_\m k_\n=m^2
\ee
where $k_\m\equiv \nabla_\m\psi_0$. The fact that the mass is constant implies that the motion is geodesic
\be
\dot{k}_\m\equiv k^\a \nabla_\a k_\m= k^\a \nabla_\m k_\a=0
\ee
The fields in this approximation reduce to classical particles moving on geodesics so that is the content of geometrical optics. To second order (physical optics), $O(1/\e)$ we have
\be
\nabla_\m k^\m + 2 k_\m p^\m=0
\ee
where $p_\m\equiv \nabla_\m\psi_1$ and can be interpreted as determining $p$ in terms of $k$.
\par
Were we to consider now a transverse theory in which kinetic and potential terms do not couple to the metric in the same way  
\be
S_m\equiv \int \frac{1}{2}\left(f_k(g)g^{\m\n}\pd_\m\psi\pd_\n\psi-f_v(g)\hat{m}^2 \psi^2\right)
\ee
the equation of motion would have been
\be
\pd_\m\left(f_k(g)g^{\m\n}\pd_\n\psi\right)+f_v(g)\hat{m}^2\psi=0
\ee
It is not difficult to check that the geometrical optics approximation is still given by
\be
f_k(g) k^2= f_v(g) m^2
\ee
The physical optics approximation does indeed change and yields
\be
k\cdot p=-\frac{\pd_\m\left(f_k(g)k^\m\right)}{2 f_v(g)}
\ee
It is clear then that when the two arbitrary functions are equal as in (\ref{scalaraction}),
\be
f_k(g)=f_v(g)\equiv f_m(g)
\ee
the trajectories of the $\psi$-particles are geodesics, so that we can safely say that the passive gravitational
mass is the same as the inertial mass, which we take to be the same as the mass parameter in the Lagrangian:
\be
m_p=m_i\equiv m
\ee
This is exactly as if this Lagrangian were minimally coupled to the gravitational field, i.e., the GR case.
\par
It seems then that in order not to contradict the WEP a {\em sufficient} condition is the universal coupling to one and only metric as in the model considered, but the coupling does not have to be {\em minimal} in the sense that it is not Diff invariant. The situation is different from scalar-tensor where it is assumed that a frame exists in which the matter is universally and minimally coupled to certain metric.\\ 
\par
On the other hand, to compute the form of the ``active gravitational mass" in transverse theories we could try to analyze the two-body problem in the transverse context; a non-trivial task. We have found it advisable instead
 to go beyond the particle approximation and to study the case of a perfect fluid. We have not attempted the corresponding exercise in the particle approximation. In passing we will reobtain the preceding result concerning equality of inertial and passive gravitational masses so that all our argumentation is done in the same physical regime.
Suppose that one defines the fluid energy density, pressure and velocity as the quantities \cite{Mukhanov}
\bea
p&\equiv&\frac{1}{2}g^{\m\n}\pd_\m\psi\pd_\n\psi-V(\psi)\nonumber\\
\r&\equiv&\frac{1}{2}g^{\m\n}\pd_\m\psi\pd_\n\psi+V(\psi)\nonumber\\
u^\m&\equiv&\frac{g^{\m\n}\pd_\n\psi}{\sqrt{g^{\m\n}\pd_\m\psi\pd_\n\psi}}
\eea
then the energy-momentum tensor corresponding to the scalar in (\ref{scalaraction}) takes the form of that of a transverse perfect fluid. Nevertheless the tensor is not conserved automatically due to the lack of full Diff invariance. Now, the equation of motion for the scalar governs the dynamics of the fluid and in our particular case it is
\be
\pd_\m\left(f_m(g)g^{\m\n}\pd_\n\psi\right)+f_m(g)V'(\psi)=0
\ee
which can be rewritten
\be
\nabla^2\psi+V'(\psi)+g^{\m\n}\pd_\m\psi\pd_\n\chi=0
\ee
where we have defined the {\em transverse scalar} $\chi\equiv\log{\frac{f_m(g)}{\sqrt{g}}}$. Multiplying by $\pd_\n\psi$ and in terms of the quantities written above it takes the form
\be
\frac{1}{2}u_\n \left(\dot{\r}+\dot{p}\right)+\left(\r+p\right)u_\n\theta+\frac{1}{2}\nabla_\n\left(\r-p\right)+\left(\r+p\right)u_\n \dot{\chi}=0
\ee
with the following notation. A dot over a quantity means the derivative in the direction of $u^\m$, i.e.,
\be
\dot{f}\equiv u^\m \nabla_\m f
\ee
and the {\em optical expansion} of the timelike congruence \cite{Hawking} is
\be
\theta\equiv \nabla_\a u^\a
\ee
Finally,
\be
\dot{u}^\m\equiv u^\a\nabla_a u^\m
\ee
so that whenever $\dot{u}^\m=0$ the congruence is a geodesic one. In the preceding equation we have also used 
\be
\nabla_\n V(\psi)=V'(\psi)\nabla_\n\psi=\frac{1}{2}\nabla_\n\left(\r-p\right).
\ee
Taking into account that 
\bea
\nabla_\n p&=&\nabla^\m\psi\nabla_\n\nabla_\m\psi-V'\nabla_\n\psi
=\nabla^\m\psi\nabla_\m\nabla_\n\psi-V'\nabla_\n\psi\nonumber\\
&=&\frac{1}{2}u_\n \left(\dot{\r}+\dot{p}\right)+\left(\r+p\right)\dot{u}_\n-\frac{1}{2}\nabla_\n\left(\r-p\right)
\eea
and substituting we arrive to
\be
u_\n \left(\dot{\r}+\dot{p}\right)+\left(\r+p\right)u_\n\theta+\left(\r+p\right) \dot{u}_\n-\nabla_\n p+\left(\r+p\right)u_\n \dot{\chi}=0.
\ee
From this equation one can derive the usual components longitudinal and transverse with respect to the velocity. Projecting with $u^\n$ we get
\be\label{conservation}
\dot{\r}+\left(\r+p\right)\left(\theta+\dot{\chi}\right)=0
\ee
Notice that the last term is absent in GR. Owing to it, pressureless matter does not behave as
\be
\rho\sim l^{-3}
\ee
(where the length scale $l$ stems from $\theta\equiv 3\frac{\dot{l}}{l})$ nor radiation ($p=\frac{1}{3}\rho$) as
\be
\rho_r\sim l^{-4}
\ee
as is the case in standard cosmology, where $l\equiv a(t)$. Instead, in transverse gravity and for a fluid verifying the equation of state $p=\omega\r$ the corresponding relationship is
\be
\r\,l^{3\left(1+\omega\right)}=\left(\frac{\sqrt{g}}{f_m(g)}\right)^{1+\omega}
\ee
since the redefined quantity 
\be\label{redef}
\r'=\r\,e^{\left(1+\omega\right)\chi}
\ee
verifies the same continuity equation that $\r$ verifies in GR
\be
\dot{\r}'+\left(1+\omega\right)\r'\,\theta=0
\ee
This means that there is a net inflow of energy in a comoving volume; the amount of momentum nonconservation is dictated by the measure of full Diff violation, as embodied in the ratio $\frac{f_m(g)}{\sqrt{|g|}}$. This physical effect would eventually lead to a violation of Newton's third law, that is, the equality of active and passive gravitational masses.
\par
On the other hand, the transverse equation obtained projecting with the adequate projector, $h^{\s\n}\equiv g^{\s\n}-u^\s u^\n$ reads
\be\label{eom}
\left(\r+p\right)u^\m\nabla_\m u^\s=h^{\s\n}\nabla_\n p 
\ee 
which gives the acceleration $a^\s=u^\m\nabla_\m u^\s$ and is {\em exactly the same equation that in GR}. This shows that, as was the case in the particle approximation, the equations of motion are equal to the corresponding general relativistic ones.
\par  
It is worth remarking that the flow lines for pressureless matter are geodesic, i.e.,
\be
\dot{u}^\a=0
\ee
The fact that no energy density appears in this equation is a signal that there is a cancellation between passive and inertial masses, exactly as was the case in the previously studied geometrical optics limit. It is nevertheless true that in Newtonian mechanics one needs the continuity equation (that we fail to have at least at the covariant level, see (\ref{conservation})) in order to obtain equality of inertial and passive masses for a finite volume particle \cite{Will}\footnote{The argument proceeds as follows. In the Newtonian regime, we define Center of Mass quantities for the chunks of fluid (particles with a finite volume) by integrating over a finite volume with the rest mass density as a weight:
\bea
\vec{R}_{cm}&\equiv&\frac{1}{\int d^{3}\vec{x}\,\rho(\vec{x},t)}\int d^{3}\vec{x}\,\,\rho(\vec{x},t)\,\,\vec{r}(\vec{x},t)\nonumber\\
\vec{V}_{cm}&\equiv&\frac{d\vec{R}_{cm}}{dt}=\frac{1}{\int d^{3}\vec{x}\,\rho(\vec{x},t)}\int d^{3}\vec{x}\,\,\rho(\vec{x},t)\,\,\frac{d}{dt}\vec{r}(\vec{x},t)\equiv\frac{1}{\int d^{3}\vec{x}\,\rho(\vec{x},t)}\int d^{3}\vec{x}\,\,\rho(\vec{x},t)\,\,\vec{v}(\vec{x},t)\nonumber\\
\vec{A}_{cm}&\equiv&\frac{d\vec{V}_{cm}}{dt}=\frac{1}{\int d^{3}\vec{x}\,\rho(\vec{x},t)}\int d^{3}\vec{x}\,\,\rho(\vec{x},t)\,\,\frac{d}{dt}\vec{v}(\vec{x},t)\equiv\frac{1}{\int d^{3}\vec{x}\,\rho(\vec{x},t)}\int d^{3}\vec{x}\,\,\rho(\vec{x},t)\,\,\vec{a}(\vec{x},t)
\eea
where $\vec{r}$, $\vec{v}$ and $\vec{a}$ are the position, speed, and acceleration fields of the fluid elements, labelled by the pair $(\vec{x},t)$. Then, starting from the equation of motion
\be
\vec{a}=-\vec{\nabla}U 
\ee
one can integrate spatially both sides of the equation with the mass density weight and, by virtue of the continuity equation
\be
\dot{\rho}+\rho\vec{\nabla}\cdot\vec{v}=0
\ee
arrive to
\be
\vec{A}_{cm}\int d^{3}\vec{x}\rho(\vec{x},t)=\int d^{3}\vec{x}\rho(\vec{x},t)\vec{a}(\vec{x},t)=-\vec{\nabla}U(\vec{R}_{cm},t)\int d^{3}\vec{x}\rho(\vec{x},t) 
\ee
where we have also neglected the differences in Newtonian potential throughout the volume of the particle. Therefore, we arrive to the equality of inertial and passive masses with the common value $\int d^{3}\vec{x}\rho(\vec{x},t)$.}. In passing, it is in this Newtonian regime where these masses are truly well defined quantities. In this respect, the quantity $\rho'$ just defined would behave in a better way. However, these last considerations depend on the volume of the particle and, in fact, in the point particle limit they become irrelevant.\\  
\par
In order to compute the active gravitational mass, that is, the source of the gravitational field, we have to study a completely different question, namely the gravitational field produced by the $\psi$-particles as determined by gravitational equations of motion. Let us call {\em active energy momentum tensor} the second member of Einstein's equations, i.e., the source of the gravitational field. For the simple model (\ref{scalaraction}) we are considering
\be\label{tmunu}
\frac{1}{\sqrt{|g|}}T^a_{\m\n}=\frac{f_m}{\sqrt{|g|}}\frac{\d L_m}{\d g^{\m\n}}
- \sqrt{|g|} f^{\prime}_m L_m g_{\m\n}
\ee
whereas the energy-momentum tensor the matter would have enjoyed, were its coupling to the gravitational field the standard one in General Relativity, would have been
\be\label{gr}
\frac{1}{\sqrt{|g|}}T^{a,GR}_{\m\n}=\frac{1}{\sqrt{|g|}}T^{p,GR}_{\m\n}=\frac{\d L_m}{\d g^{\m\n}}
- \frac{1}{2} L_m g_{\m\n}
\ee
It is important to realize that in the case of GR this tensor has a very clear physical meaning (absent for transverse matter), in the sense that it reduces in flat space to the Noether charge associated to translational invariance, which represents in turn the energy content of the inertial mass. In somewhat pedantic terms, the Rosenfeld tensor reduces in flat space to the Belinfante one (cf. the appendix for some discussion of this point.)
\par
Let us examine this energy-momentum sources in the fluid approximation and restricting ourselves again to only one species of matter field, $\psi$, for simplicity. In terms of the constructs defined above the energy-momentum tensor is
\be\label{emtensor}
\frac{2}{\sqrt{|g|}}T^a_{\m\n}=\frac{f_m}{\sqrt{|g|}}(\rho+p)u_\m u_\n
- 2\sqrt{|g|} f^{\prime}_m\, p\, g_{\m\n}
\ee
and similarly in the case of GR we have
\be
\frac{2}{\sqrt{|g|}}T^{a,GR}_{\m\n}=(\rho+p)u_\m u_\n
-  p g_{\m\n}
\ee
The scalar source of gravitation, i.e., the {\em active gravitational mass} is then
\be
m_a^{GR}\equiv T_{\a\b}u^\a u^\b= \rho
\ee
In transverse theories this is no longer the case. The corresponding quantity is
\be
m_a=\frac{f_m}{\sqrt{|g|}}\rho +\left(\frac{f_m}{\sqrt{|g|}}- 2\sqrt{|g|} f^{\prime}_m\right) p
\ee
We can measure the relative difference between general relativistic and transverse active masses through the lowest order quantity
\be\label{delta}
\d\equiv \frac{m_a-m_a^{GR} }{m_a^{GR}}=\left(\frac{p}{\rho}\frac{f_m-2g f_m^\prime}{\sqrt{|g|}}+\frac{f_m-\sqrt{|g|}}{\sqrt{|g|}}\right)
\ee
In the nonrelativistic {\em cold} limit we can approximate
\be
\frac{p}{\rho}\sim 0
\ee
but even in that case the second member produces a difference between the active masses of both theories. In general relativity the three masses are equal, whereas active masses are different in both theories. It is worth pointing out that, even for a fluid with equation of state $p=\omega\r$ and redefining the energy density as in (\ref{redef}) (so that we get a conservation of energy equation like the one in GR), it is clear that the equation of motion (\ref{eom}) and the energy momentum-tensor (\ref{emtensor}) do not take the GR form with $\r'$ replacing $\r$, except when $\omega=0$. If we admit that in GR all three masses are equal and that all the content of the theory is embodied in the equations considered, then one must conclude that all three masses cannot be equal in the transverse model we are discussing. As we have mentioned, this causes either a violation of the WEP or more likely a violation of Newton's third law which must be carefully tuned up in order for it to be compatible with experiment.

\section{Comments on the Newtonian limit}
Let us study the Newtonian limit of the particular class of transverse theories considered in this paper from the complementary viewpoint of the point-particle action
\footnote{It is perhaps worth remarking that in contradistinction with the eikonal limit of field theory mentioned in the main text, the equations of motion are not just geodesics. We would like to avoid here the consideration of energy momentum tensors with support on the world line of the point particle (cf. \cite{Poisson} for a detailed review).} following the well-known Landau--Lifshitz approach \cite{Landau}. The natural starting point is
\be\label{accion}
S_T\equiv-m_T c\int f\left(g\right) ds
\ee
Where $f(g)\equiv \frac{f_m}{\sqrt{|g|}}$. Its weak field, low velocities limit is given by
\be
S_T=-m_T c^2 \int f\left(1+\kappa|h|\right)\sqrt{1+\kappa h_{00}-\frac{\vec{v}^2}{c^2}}dt
\ee
where we have also expanded $g_{\m\n}=\eta_{\m\n}+\kappa h_{\m\n}$. It is a fact that $h=h_{00}+\hat{h}$ with $\hat{h}\equiv\sum_i h_i^i$.
The nonrelativistic action for a particle propagating in a Newtonian gravitational field characterized by a potential energy per unit mass $\Phi_N$, , i.e.
\be
S_{NR}=-m\int dt\left(c^2-\frac{\vec{v}^2}{2}+\Phi_N+\ldots\right)
\ee 
Enforcing $S_T=S_{NR}$ leads to
\bea
m&=&m_T f(1)\nonumber\\
\kappa h_{00}&=&\frac{2 f(1)}{\left(f(1)+f^\prime(1)\right)c^2}\left(\Phi_N-\kappa c^2\frac{f^\prime(1)}{f(1)}\hat{h}\right)
\eea
In order for this to be a consistent expansion we have to assume that $\frac{f^\prime(1)}{f(1)}$ is also a small number.
The transverse energy momentum tensor is given by (\ref{tmunu})
\be
T_{00}=f(g)\rho_T c^2=\left(f(1)+\kappa f^\prime(1)h+O(\kappa^2)\right)\rho_T c^2
\ee
and the Ricci tensor in that limit is
\be
R_{00}\sim \sum_i\pd_i \Gamma^i_{00}\sim \frac{1}{2}\sum_i\pd_i \pd^i g_{00}\sim \frac{1}{2}\Delta \left(\kappa h_{00}\right)
\ee
The leading term in Einstein's equations 
\[R_{00}=\frac{c \kappa^2}{2}T_{00}\]
(where we have defined, following Landau and Lifshitz, $T_{\m\n}\equiv \frac{2 c}{\sqrt{|g|}}\frac{\d S_m}{\d g^{\m\n}}$)
is then
\be
\Delta \Phi_N\sim \frac{f(1)}{2}c^3 \kappa^2 \rho_T
\ee
  Poisson's equation is eventually recovered provided\footnote{
  One could be tempted to identify
  \[\kappa^2\equiv \frac{8\pi G}{f(1) c^3}\] while leaving $\rho=\rho_T$. This is fine for recovering Poisson's equation, but then the particle action (\ref{accion}) does not enjoy the proper limit. This is the reason why we prefer the identification as in (\ref{texto}).  
  }
\bea\label{texto}
m&=&m_T f(1)\nonumber\\
\kappa^2&\equiv&\frac{8\pi G}{c^3}\nonumber\\
\rho&=&\rho_T f(1)
\eea

\section{Conclusions}

In this paper it has been shown that gravitational theories in which the purely gravitational action is invariant only under the transverse subgroup of diffeomorphisms are subject to the usual constraints on scalar-tensor theories since they propagate an additional scalar degree of freedom in the metric. Bounds on the form of the arbitrary functions one is free to add are discussed in detail in the literature and are straightforward to apply to transverse models. The scalar is not postulated but appears automatically due to the lack of symmetry. In this sense, it is a consequence of the spacetime symmetry principle. 
\par
On the other hand, when the gravitational action is Einstein--Hilbert (so that the scalar no longer propagates) but the matter part is only transverse invariant, it is predicted that the three different concepts of mass one can discuss cannot be equal as is the case in GR. In the geometrical optics limit of a particular class of models it can be found that inertial and passive gravitational masses are equal between them and to the general relativistic values. Therefore, one is tempted to conclude that active a passive gravitational masses are indeed different
\be
m_a\neq m_p
\ee
which violates Newton's third law, i.e., momentum conservation. Computations in the fluid approximation seem to support this conclusion since one can derive an equation of motion for the flow lines that is the same we encounter in GR. This conclusion is nonetheless delicate, since it is by no means easy to define concepts of mass in covariant theories. In fact, when the equation of state of the fluid can be written as $p=\omega\r$ and redefining the energy density as in (\ref{redef}), one gets for the redefined quantity $\r'$ a conservation of energy equation like the one in GR. However, neither the equation of motion nor the energy momentum tensor for $\r'$ take the GR form. In view of this and taking into account the clues given by the geometrical optics approximation, we prefer to give all physical effects in terms of the original quantities. Comparison with the GR template is easier, we do not need any assumption on the equation of state of the fluid and all observable effects are clearly embodied in the energy-momentum non-conservation parameter $\d$.
\par
It is not clear to what extent these theories are constrained by existing observations.  The observable (\ref{delta}) depends on the spacetime point; this means that energy would weigh differently according to the particular position in spacetime of the object considered. The fact that in the models we have analyzed this effect is {\em universal}, i.e., the same for all bodies, will make it very difficult to check observationally.
\par
Current experimental bounds for this violation; that is, for the {\em differential} violation in different bodies, $\left|\left(\frac{m_a}{m_p}\right)_1-\left(\frac{m_a}{m_p}\right)_2\right|$ are of order $\leq 10^{-13}$, so that this is the maximum acceptable order of magnitude of the appropriate differences of the observable we have used to parameterize transverse deviations from general relativity
\be
\Delta \d\equiv\d_1-\d_2\leq 10^{-13}
\ee
\par
But this does not constrain the observable $\d$ itself. It is nevertheless true that even if units are chosen in such a way that $m_a= m_p$ at some point in the spacetime, this ratio would change with cosmic history following the variation of the metric. As we have pointed out, the ratio between both masses is Newton's constant so the existing limits on the ``constancy of constants" should apply. The limit on the relative variation of Newton's constant
\be
\frac{\dot G}{G}\leq 10^{-12}year^{-1}
\ee
which comes from helioseismology \cite{Krauss} then puts a similar constraint on $\d$. Notice that generically a theory that predicts a coupling constant variable with the spacetime point also suffers from violations of Newton's third law.
\par
Another characteristic generic prediction of transverse gravity theories is that of {\em dipolar} gravitational radiation.
The standard argument as to the reason of why the dipolar component vanishes in General Relativity is based on momentum conservation, a luxury we can not afford in transverse theories. In fact, it has been conjectured \cite{Will} that a theory of gravity predicts no dipolar gravitational radiation if and only if it satisfies the {\em Strong Equivalence Principle} (SEP). Given that in our case inequality between active and passive masses implies a local gravitational constant then violation of the SEP is not surprising. Dipolar radiation should be proportional, at first order, to the parameter that measures the violation of momentum conservation, that is, $\Delta\d$. A possible dipolar component of radiation is therefore suppressed at least by a factor $10^{-12}$.
\par
Yet another consequence of the presence of the arbitrary $g(x)$ function in the matter Lagrangian will be the violation of Local Position Invariance (LPI) \cite{Will}. Consider for example the effect on the Electromagnetic Lagrangian: there will be a varying dielectric constant and hence a varying $\a_{EM}$. Therefore,  constrains on the variability of all fundamental constants (not only the gravitational $G$) will apply in general. For a comprehensive review on the bounds over these see \cite{Uzan}.
\par
Once more, it is indeed remarkable how tight is Einstein's physical scheme incorporating the gravitational field. The restriction of the invariance of General Relativity one makes in transverse theories is a minimal one, in the sense that TDiff(M) is the {\em maximal} subgroup of the full Diff(M) group. In the flat case it includes, in particular, the full set of Lorentz transformations, so the transverse condition poses no restriction at all. In spite of all this, the experimentally allowed window to make a transverse deformation of General Relativity is \footnote{And this in a particular class of transverse models, which are {\em a priori} the ones that are expected to be closest to General Relativity.} quite small, as we have argued in the main body of the paper.

\section*{Acknowledgements}
 We are grateful to Eduard Mass\'o and Enric Verdaguer for useful discussions. This work has been partially supported by the
European Commission (HPRN-CT-200-00148) and by FPA2006-05423 (DGI del MCyT, Spain) ,
Proyecto HEPHACOS ; P-ESP-00346 (CAM) and Consolider PAU, CSD-2007-00060.        

\appendix
\section{Transverse energy-momentum tensors and their conservation laws.}
There are indeed several energy momentum tensors of interest in the transverse case. Actually, they are in 
general not true tensors under GR but only densities.
\par
 The {\em active energy-momentum} tensor (that is, the source of the gravitational equations), is
\be
T_{\m\n}\equiv\frac{\d S_m}{\d g^{\m\n}}\equiv \frac{\d}{\d g^{\m\n}}\int d^n x f_m L_m
\ee

 In order to study the transverse conservation law, let us perform a TDiff,
under which
\be
\d g_{\a\b}=\pounds(\xi)g_{\a\b}\equiv \xi^{\rho}\pd_\rho g_{\a\b}+g_{\a\rho}\pd_\b \xi^\rho 
+ g_{\rho\b}\pd_\a \xi^\rho
\ee
The use of covariant derivatives is best avoided for the time being. The fact that the
quantities considered are not tensors under Diff has already been mentioned, and this can 
obscure the reasoning. Performing a TDiff on the matter action 
\bea
0&=&\d_{T-diff}S_m=\int d^n x \left(\e^{\rho\m_2\ldots\m_n}\pd_{\m_2}\Omega_{\m_3
\ldots \m_n}\pd_{\rho} g_{\a\b}+
g_{\a\rho}\pd_\b\left(\e^{\rho\m_2\ldots\m_n}\pd_{\m_2}\Omega_{\m_3
\ldots \m_n}\right)\right. \nonumber\\
&&\left.+g_{\b\rho}\pd_\a\left(\e^{\rho\m_2\ldots\m_n}\pd_{\m_2}\Omega_{\m_3
\ldots \m_n}\right)\right) T^{\a\b}
\eea
Taking into account that $\e^{\m_1\ldots\m_n}$ is independent of the metric, and denoting
\be
\omega^{\m\n}\equiv \e^{\m\n\m_3\ldots\m_n}\Omega_{\m_3\ldots\m_n}
\ee
the aforementioned condition is equivalent to:
\be
0=\int d^n x\,\omega^{\m\n}\left(-\pd_\m g_{\a\b}\pd_\n 
T^{\a\b}+2\pd_\n \pd_\l T_\m\,^\l\right) 
\ee
This means that
\be\label{cons}
\pd_\m g_{\a\b}\pd_\n T^{\a\b}-\pd_\n g_{\a\b}\pd_\m T^{\a\b}=
2\left(\pd_\n \pd_\l T_\m\,^\l-\pd_\m \pd_\l T_\n\,^\l\right)
\ee
which does imply 
\be\label{fi}
\pd_\l T_\m\,^\l-\frac{1}{2}\pd_\m g_{\a\b}T^{\a\b}=\pd_\m \Lambda
\ee
where $\Lambda$ is an arbitrary function. 
Using the well-known formula (valid for any symmetric tensor)

\be
\nabla_\n S_\m\,^\n=\frac{1}{\sqrt{|g|}}\pd_\n\left(\sqrt{|g|}S_\m\,^\n\right)-\frac{1}{2}
\pd_\m g_{\a\b}S^{\a\b}
\ee
this can be rewritten as
\be\label{bianchi}
\nabla_\n\left(\frac{T_\m\,^\n}{\sqrt{|g|}}\right)=\frac{1}{\sqrt{|g|}}\,\,\pd_\m\Lambda
\ee
in the understanding that the covariant derivative is to be taken as if  $T_{\a\b}$ were a 
true tensor. Physically this means that the active energy-momentum tensor is neither conserved nor covariantly conserved.
On the other hand we recall that
\be
T^a_{\m\n}=f_m\frac{\d L_m}{\d g^{\m\n}}
- g f^{\prime}_m L_m g_{\m\n}
\ee
We have used the abbreviation $\frac{\d L_m}{\d g^{\m\n}}$ instead of the most accurate 
$\frac{\d \int d^n x L_m}{\d g^{\m\n}}$. It is interesting to study the nature of this tensor
since as we have seen it is not possible to deduce its covariant conservation using only 
invariance under TDiffs.\footnote{ Let us now analyze the invariance of a Diff invariant theory considering longitudinal and transverse Diffs separately. It is a fact of life that any Diff can be written (perhaps under certain global conditions) as
\be
\xi^\m=\xi^\m_L+\xi^\m_T
\ee
where
\bea
\xi^\m_T&=&\pd_\rho \omega^{\m\rho}\nonumber\\
\xi^\m_L&\equiv& \pd^\m \Phi
\eea
with $\omega^{(\m\rho)}=0$.

It is amusing to do the counting of the transverse part in detail: in terms of forms we are representing
$\xi_1=\d\omega_2$, but this $\omega_2$ is defined up to a $d\omega_3$ which in turn is defined up to a total differential
and so on. When the dust settles down, we get
\be
(1-1)^n-1+n=n-1
\ee
Invariance under longitudinal Diffs just means that
\bea
0&=&\int d^n x \,\,\pd^\rho \Phi\,\,\left(T^{\a\b}\pd_\rho g_{\a\b}-\pd_\b\left(T^{\a\b}g_{\a\rho}\right)-\pd_\a\left(T^{\a\b}g_{\rho\b}\right)\right)=\nonumber\\
&&\int d^n x \,\,\pd^\rho \Phi\,\,\left(T^{\a\b}\pd_\rho g_{\a\b}-2\pd_\a T^\a_\rho\right)=\nonumber\\
&&-\int d^n x \,\,\Phi\,\,\pd^\rho\left(T^{\a\b}\pd_\rho g_{\a\b}-2\pd_\a T^\a_\rho\right) 
\eea
This implies that
\be
\pd^\rho\left(T^{\a\b}\pd_\rho g_{\a\b}-2\pd_\a T^\a_\rho\right)=0 
\ee
But we knew already that the transverse invariance enforces that
\be
\pd_\l T_\m\,^\l-\frac{1}{2}\pd_\m g_{\a\b}T^{\a\b}=\pd_\m \Lambda
\ee
so the two together yield
\be
\Box \Lambda=0
\ee
which gives 
\be
\Lambda=0
\ee
if the correct boundary conditions are used.
}
\section{Conservation of the active gravitational mass}
 Let us recall the usual definition of energy. Given a timelike Killing vector, we define
\be
j^\a\equiv T^\a_\m k^\m
\ee
so that, owing to the Killingness,
\be
\nabla_\a j^\a=k^\m\nabla_\a T_\m^\a
\ee
We now define the one-form
\be
j\equiv j_\m dx^\m
\ee
and apply Stokes to the cylindrical region $C$ surrounding the flow lines of the Killing at an spatial distance
$R\rightarrow\infty$ and with two caps at two values of the parameter $\l=0$ and $\l=T$ (remember that $k=\frac{\pd}{\pd t}$
in flat space). Then
\bea
\int_C \sqrt{|g|}\nabla_\a j^\a d^n x&=&\int_C d*j=\int_{\pd C}j=\frac{1}{(n-1)!}\int_{\pd C}\sqrt{|g|}j^\a \e_{\a\m_1\ldots\m_{n-1}}dx^{\m_1}\wedge\ldots \wedge dx^{\m_{n-1}}\nonumber\\
&=&E(T)-E(0)
\eea
where the energy corresponding to the value $\l$ of the Killing flow parameter is defined as
\be
E(\l)=\frac{1}{(n-1)!}\int_{\l}\,\,\sqrt{|g|}\,\,\frac{k\cdot j}{k^2}\,k^\a \e_{\a\m_1\ldots\m_{n-1}}dx_T^{\m_1}\wedge\ldots \wedge dx_T^{\m_{n-1}}
\ee
(where we have divided the spacetime coordinates, into the Killing parameter, $\l$ on the one hand, and the transverse coordinates $x_T$ on the other) and the equality above holds provided $j \rightarrow 0$ fast enough when $R\rightarrow\infty$.
 \par
 In GR the energy-momentum is conserved, which in turn implies that
 \be
 E(T)=E(0)
 \ee
To summarize, in transverse theories, with asymptotically flat boundary conditions, we do not have conservation of energy, but rather
 \be
 E(T)=E(0)+\int_C k^\m \pd_\m\Phi
 \ee
If the scalar $\Phi$ is also invariant under the transformation generated by the Killing,
\be
\pounds(k)\Phi\equiv k^\m \pd_\m \Phi=0
\ee
or even if $\Phi(\l=T,x_T)=\Phi(\l=0,x_T)$, then the Rosenfeld energy is still conserved. This seems then the most natural setup when the spacetime enjoys Killing symmetries.
\section{Belinfante versus Rosenfeld}
The label {\em energy-momentum tensor} for the above construct, the active energy-momentum tensor (\ref{tmunu}) can indeed be 
questioned for
very good reasons. It is a metric (Rosenfeld) tensor which is not conserved, and consequently,
it does not reduce in flat space to the canonical one, or to its equivalent Belinfante form. That is, the tensor
(\ref{tmunu}) does not convey the Noether current corresponding to translation invariance.
In order to illustrate this, let us consider the simplest example, namely a real scalar field
without coupling to the determinant of the metric, i.e., $f_m(g)=1$
\be
S_m\equiv\int d^n x L_m =\int d^n x \frac{1}{2} g^{\m\n}\pd_\m\phi\pd_\n\phi
\ee 
The energy-momentum tensor as defined before is
\be
T_{\m\n}=\frac{1}{2}\pd_\m\phi\pd_\n\phi
\ee
Using the equation of motion (EM) of the scalar
\be
\frac{\d S_m}{\d \phi}\equiv\pd_\m\left(g^{\m\n}\pd_\n\phi\right)=0
\ee
it can be shown that
\be\label{conser}
\sqrt{|g|}\nabla_\n\left(\frac{T_\m\,^\n}{\sqrt{|g|}}\right)=\frac{1}{2}\nabla_\m L_m
\ee
conveying that fact that this energy momentum is not covariantly conserved in the general case, and
thus it cannot act as a consistent source of Einstein's equations if the gravitational part is Einstein--Hilbert. 
What is worse, $T_{\m\n}$ does not reduce in flat
space to the canonical one
\be
T_{\m\n}^{can}=\pd_\m\phi\pd_\n\phi-\frac{1}{2}L_m \eta_{\m\n}
\ee
which is well known to be conserved.
This does not happen of course with the usual covariant Lagrangian
\be
S_{cov}=\int d^n x \sqrt{|g|}\frac{1}{2}g^{\m\n}\pd_\m\phi\pd_\n\phi
\ee
whose energy-momentum tensor
\be
T_{\m\n}^{GR}\equiv\frac{2}{\sqrt{|g|}}\left(\frac{1}{2}\sqrt{|g|}\pd_\m\phi\pd_\n\phi-
\frac{1}{4}\sqrt{|g|}g_{\m\n}g^{\a\b}\pd_\a\phi\pd_\b\phi\right)
\ee
is both covariantly conserved thanks to the new EM and reduces to the canonical one in flat space.


\end{document}